\documentclass[final,5p,times,twocolumn]{elsarticle}
\journal{Phys. Lett. B}

\usepackage{graphicx,color}
\usepackage{amsmath,amssymb}
\usepackage{dcolumn}

\newcommand{\eq}[1]{\begin{equation}#1\end{equation}}

\newcommand{\ket}[1]{\ensuremath{\,|{#1}\rangle}}

\newcommand{\expect}[1]{\ensuremath{\langle{#1}\rangle}}

\newcommand{\op}[1]{\ensuremath{#1}}

\renewcommand{\vec}[1]{\ensuremath{\mathbf{#1}}}

\newcommand{\HO}{\ensuremath{\op{H}}}

\newcommand{\TO}{\ensuremath{\op{T}}}

\newcommand{\VO}{\ensuremath{\op{V}}}

\newcommand{\pOV}{\ensuremath{\vec{\op{p}}}}

\newcommand{\POV}{\ensuremath{\vec{\op{P}}}}

\newcommand{\XOV}{\ensuremath{\vec{\op{X}}}}

\newcommand{\UCOM}{\ensuremath{\textrm{UCOM}}}
\newcommand{\intr}{\ensuremath{\textrm{int}}}

\newcommand{\cm}{\ensuremath{\textrm{cm}}}
\newcommand{\elem}[2]{\ensuremath{{}^{#2}\text{#1}}}

\newcommand{\symboldiamond}[1][black]{{\color{#1}$\blacklozenge$}}

\newcommand{\symbolcircle}[1][black]{{\color{#1}$\bullet$}}

\definecolor{FGViolet}{rgb}{0.61,0.32,0.61}
\definecolor{FGDarkBlue}{rgb}{0,0,0.6}
\definecolor{FGBlue}{rgb}{0,0,0.8}
\definecolor{FGLightBlue}{rgb}{0.2, 0.6, 0.8}
\definecolor{FGGreen}{rgb}{0.2,0.7,0.2}
\definecolor{FGLightGreen}{rgb}{0.4,1,0.4}
\definecolor{FGYellow}{rgb}{1,0.95,0}
\definecolor{FGOrange}{rgb}{0.95,0.5,0.1}
\definecolor{FGRed}{rgb}{0.8,0,0}
\definecolor{FGWhite}{rgb}{1,1,1}
\definecolor{FGLightGray}{rgb}{0.8,0.8,0.8}
\definecolor{FGGray}{rgb}{0.5,0.5,0.5}
\definecolor{FGDarkGray}{rgb}{0.3,0.3,0.3}
\definecolor{FGBlack}{rgb}{0,0,0}

\begin{document}

\begin{frontmatter}

\title{Center-of-mass problem in truncated configuration interaction and coupled-cluster calculations}

\author[tud]{Robert Roth}
\ead{robert.roth@physik.tu-darmstadt.de}

\address[tud]{Institut f\"ur Kernphysik, Technische Universit\"at Darmstadt,
64289 Darmstadt, Germany}

\author[msuc]{Jeffrey R. Gour}

\author[msuc,msup]{Piotr Piecuch}
\ead{piecuch@chemistry.msu.edu}

\address[msuc]{Department of Chemistry, Michigan State University, East Lansing, 
MI 48824-1322, USA}

\address[msup]{Department of Physics and Astronomy, Michigan State University, East Lansing, 
MI 48824-1322, USA}

\begin{abstract}  
The problem of center-of-mass (CM) contaminations in \emph{ab initio} nuclear structure calculations using configuration interaction (CI) and coupled-cluster (CC) approaches is analyzed. A rigorous and quantitative scheme for diagnosing the CM contamination of intrinsic observables is proposed and applied to ground-state calculations for \elem{He}{4} and \elem{O}{16}. The CI and CC calculations for \elem{O}{16} based on model spaces defined via a truncation of the single-particle basis lead to sizable CM contaminations, while the importance-truncated no-core shell model based on the $N_{\max}\hbar\Omega$ space is virtually free of CM contaminations.
\end{abstract}


\end{frontmatter}

\section{Introduction}

In the realm of quantum many-body systems the atomic nucleus poses unique challenges that do not appear in the theoretical description of other systems. One of them results from the fact that the nucleus is a finite self-bound system. Unlike, e.g., the many-electron systems in atomic, molecular, and condensed matter physics, and chemistry, in which the binding originates from an external potential created by much heavier atomic nuclei that define a reference frame, the nucleus is bound solely by the interactions between the constituent nucleons without any external confinement.

In the exact theory, the many-body state $\ket{\Psi}$ of the nucleus factorizes into the intrinsic state $\ket{\psi_{\intr}}$ and a state $\ket{\psi_{\cm}}$ describing the dynamics of the center-of-mass (CM),
\eq{ \label{eq:decoupling}
  \ket{\Psi} = \ket{\psi_{\intr}} \otimes \ket{\psi_{\cm}} \;.
}
Only this complete decoupling guarantees that the intrinsic state of the nucleus is translationally invariant and, as result, that all intrinsic observables are free of spuriousities induced by the CM state of the system. The dynamics of $\ket{\psi_{\intr}}$ is governed solely by the intrinsic Hamiltonian
\eq{
  \HO_{\intr} 
  = (\TO - \TO_{\cm}) + \VO
  = \TO_{\intr} + \VO \, ,
}
where $\TO_{\intr} = \frac{1}{2mA}\sum_{i<j}^{A} (\pOV_i - \pOV_j)^2$ is the intrinsic kinetic energy and $\VO$ is the nuclear interaction.

Generally, the wave function factorization given by Eq. (\ref{eq:decoupling}) no longer holds in approximate calculations. One way to enforce it is through the use of Jacobi coordinates, as in, e.g., the translationally invariant formulation of the no-core shell model (NCSM) \cite{NaKa00,NaQu09}, the hyperspherical harmonics approach \cite{BaLe99}, and, implicitly, also the Green's function Monte Carlo method \cite{PiWi01}. Due to the computational complexity, these approaches are limited to small few-body systems. For this reason, most approaches aimed at heavier nuclei adopt some finite-dimensional model space spanned by Slater determinants constructed with a single-particle basis set, such as the harmonic oscillator (HO) basis. Generally, the individual Slater determinants are not translationally invariant and most model spaces do not restore this invariance automatically, i.e., the solution of the many-body problem for $\HO_{\intr}$ leads to states with a coupling between intrinsic and CM motions which is induced by and depends on the structure of the model space. The one exception is the $N_{\max}\hbar\Omega$ space of NCSM \cite{NaKa00,NaQu09}. For a Slater-determinant basis of HO single-particle states truncated with respect to the HO excitation energy $N\hbar\Omega$ there exists a unitary mapping onto a Jacobi-coordinate basis. Thus, the exact factorization of Eq. \eqref{eq:decoupling} is possible in the $N_{\max}\hbar\Omega$ space. For other single-particle bases or model-space truncations, the factorization is lost and a coupling of intrinsic and CM states emerges. In this communication, we provide a quantitative analysis of the CM contamination in the truncated {\it ab initio} configuration interaction (CI) and coupled-cluster (CC) calculations for \elem{He}{4} and \elem{O}{16}.

\section{Center-of-mass diagnostics}

Because a direct analysis of the CI and CC wave functions with respect to the factorization given by Eq. \eqref{eq:decoupling} is not feasible, we need a practical tool to assess the degree of unphysical coupling between intrinsic and CM motions resulting from the truncations used in the CI and CC calculations. A stringent yet simple probe can be proposed based on replacing $\HO_{\intr}$ entering the CI and CC calculations by
\eq{ \label{eq:hamiltonianmod}
  \HO_{\beta} = \HO_{\intr} + \beta\; \HO_{\cm} ,
}
which includes a Hamilton-type operator $\HO_{\cm}$ acting exclusively on the CM part of the many-body state of interest, with parameter $\beta$ controlling its strength. In principle, any operator depending on the CM position $\XOV_{\cm}$ and momentum $\POV_{\cm}$, with a spectrum bounded from below, could be used for this purpose, but we further require that the exact ground state of $\HO_{\cm}$ can be represented in the model space. Thus, in a ground-state calculation with $\HO_{\beta}$ for non-zero $\beta$ the CM is in its exact ground state if the corresponding many-body wave function $\ket{\Psi_{\beta}}$ factorizes.

For calculations based on Slater determinants of HO single-particle states, an operator which meets these requirements is
\eq{ \label{eq:cmhamiltonian}
  \HO_{\cm} 
    = \frac{1}{2mA}\POV_{\cm}^2 + \frac{mA \Omega^2}{2} \XOV_{\cm}^2 - \frac{3}{2}\hbar\Omega \;,
}
as was first adopted by Palumbo \cite{Palu67} as well as Gloeckner and Lawson \cite{GlLa74} in the context of the CM problem in the valence-space shell model \cite{RaFa90}. In an $N_{\max}\hbar\Omega$ space, the lowest $N_{\max}+1$ eigenstates (including both parities) of $\HO_{\cm}$, Eq. (\ref{eq:cmhamiltonian}), are reproduced exactly. Hence, as long as the $0\hbar\Omega$ space is a subspace of the model space used in the CI or CC calculation, the solution of the $\HO_{\cm}$ eigenvalue problem leads to the exact CM ground-state energy, which is zero by construction. For a closed-shell nucleus, the use of the $0p0h$ determinant as a reference state guarantees this.

By solving the Schr{\" o}dinger equation for $\HO_{\beta}$, Eq. (\ref{eq:hamiltonianmod}), using different $\beta$ values we can determine to what extent the intrinsic and CM components of the many-body wave function of interest are coupled. If a given many-body method leads to a factorized state $|\Psi\rangle$, as in Eq. \eqref{eq:decoupling}, the intrinsic component $\ket{\psi_{\intr}}$ and the intrinsic observables become independent of $\beta$. In this case, the $\beta \HO_{\cm}$ part of $\HO_{\beta}$ can only affect the CM component of $|\Psi\rangle$, which has no effect on intrinsic properties. However, if the many-body scheme used to solve the Schr{\" o}dinger equation for $|\Psi\rangle$ cannot factorize intrinsic and CM motions, the intrinsic observables, most notably the expectation value of $\HO_{\intr}$, will acquire an unphysical dependence on $\beta$. To quantify the strength of this unphysical coupling and its impact on the intrinsic energy, we need to monitor the expectation value $\expect{\HO_{\intr}}_{\beta}$ computed with the eigenstates $\ket{\Psi_{\beta}}$ of $\HO_{\beta}$. A necessary condition for the wave function factorization given by Eq. \eqref{eq:decoupling} is that $\expect{\HO_{\intr}}_{\beta}$ is independent of $\beta$. Thus, we adopt the change of the intrinsic energy, $\Delta\expect{\HO_{\intr}}_{\beta} = \expect{\HO_{\intr}}_{\beta} - \expect{\HO_{\intr}}_{0}$, when going from $\beta=0$ to a finite $\beta$, as a \emph{primary criterion} for assessing the CM contamination in the wave function. This criterion provides a quantitative measure of the impact of the CM contamination on the intrinsic energy, while identifying states $|\Psi\rangle$ that factorize according to Eq. \eqref{eq:decoupling} which satisfy $\Delta\expect{\HO_{\intr}}_{\beta} = 0$.

As a \emph{secondary criterion} we use the expectation value of $\HO_{\cm}$ obtained for non-zero values of $\beta$. If a given ground-state calculation allows for a factorization of the wave function, we can expect $\expect{\HO_{\cm}}_{\beta}$ to be \emph{exactly} equal to the lowest eigenvalue of $\HO_{\cm}$, i.e., zero in the case of $\HO_{\cm}$ defined by Eq. \eqref{eq:cmhamiltonian}. The non-zero $\expect{\HO_{\cm}}_{\beta}$ value at a finite $\beta$ indicates a coupling of CM and intrinsic motions. We must emphasize, however, that the use of $\expect{\HO_{\cm}}_{\beta}$ alone is treacherous, since it does not provide any quantitative measure of the effect of the coupling of the intrinsic and CM motions on intrinsic properties. The expectation value of $\HO_{\cm}$ at $\beta=0$, which is sometimes used to judge the magnitude of CM contamination in approximate many-body calculations, does not provide any quantitative information, since $\expect{\HO_{\cm}}_{\beta=0}$ can assume any positive value when the calculated wave function $|\Psi\rangle$ is factorizable.

We use $\HO_{\beta}$, Eq. \eqref{eq:hamiltonianmod}, solely for probing the spurious CM coupling at moderate $\beta$ values, although its original use as a way to eliminate the CM contamination was tied to large values of $\beta$ ($\sim 10^5$) \cite{GlLa74}. In a CI calculation, the use of $\HO_{\beta}$ with large $\beta$ amounts to a projection of the model space onto the largest embedded $N_{\max}\hbar\Omega$ space. In comparison to NCSM, this approach is inefficient and numerically problematic.

\section{Many-body methods}

We employ the above diagnostics to analyze CM contaminations in \emph{ab initio} CI and CC calculations, which rely on the model spaces spanned by Slater determinants constructed from the HO single-particle states that satisfy the condition $e \leq e_{\max}$, where $e = 2n+l$ is the principal quantum number, i.e., model spaces that violate translational invariance from the outset. The first method we consider is the importance-truncated CI (IT-CI) approach introduced in Refs. \cite{RoNa07,RoGo09,Roth09} in which the CI eigenvalue problem in the above model space is further reduced through the use of an importance measure for the individual many-body basis states derived from perturbation theory. We focus on the IT-CI approach with up to $4p4h$ excitations from the initial reference state $|\Phi\rangle$ [IT-CI($4p4h$)], which for the closed-shell nuclei examined in this work is the determinant obtained by occupying the $A$ lowest single-particle HO states. In the limit of vanishing importance threshold, IT-CI($4p4h$) becomes equivalent to the CI method with all singly, doubly, triply, and quadruply excited determinants relative to $|\Phi\rangle$ [${\rm CISDTQ \equiv CI}(4p4h)$]. In this study, we follow the recipe described in detail in Refs. \cite{RoGo09,Roth09}, in which we perform a sequence of IT-CI($4p4h$) calculations with decreasing thresholds and extrapolate the results to the vanishing importance threshold limit, thus obtaining a very good approximation to a complete CISDTQ calculation.

The IT-CI approach offers two important advantages for the present study. First, it allows for a computationally inexpensive evaluation of the exact quantum-mechanical expectation values $\expect{\HO_{\intr}}_{\beta}$ and $\expect{\HO_{\cm}}_{\beta}$ which require much less effort than the corresponding CISDTQ calculations. Second, we can apply the same importance-truncation methodology to other model space truncations, such as the $N_{\max}\hbar\Omega$ space of NCSM. In the limit of vanishing importance threshold, the resulting sequential importance-tuncation IT-NCSM(seq) scheme, which we make use of in this study as well, reproduces the model space of the full NCSM approach \cite{Roth09}.

The second many-body approach we employ is the single-reference CC theory \cite{coester,coester2,cizek,cizek2}. The specific CC models considered in this study include CCSD (CC singles and doubles) \cite{ccsd} and CR-CC(2,3) \cite{crccl}, both used in our recent nuclear structure work \cite{RoGo09,kowalski04,o16prl,ni56_2007}. In the CCSD approach, the cluster operator $T$ defining the CC ground state $|\Psi \rangle = \exp(T) |\Phi\rangle$, where in the exact case $T = T_{1} + T_{2} + \cdots + T_{A}$ and $T_{n}$ designates the $npnh$ component of $T$, is truncated at the $2p2h$ clusters $T_{2}$, i.e., $T = T_{1} + T_{2}$. The $T_{1}$ and $T_{2}$ clusters are determined by solving the system of non-linear, energy-independent equations obtained by inserting the CC wave function, in which $T = T_{1} + T_{2}$, into the Schr{\" o}dinger equation and projecting on the singly and doubly excited determinants relative to $|\Phi\rangle$. The CCSD energy is calculated afterwards using the converged $T_{1}$ and $T_{2}$ clusters. Since $T_{3}$ clusters are important to obtain a quantitative description of closed-shell systems, we correct the CCSD results for their effect through the suitably defined non-iterative corrections to CCSD energies, calculated using the CCSD values of $T_{1}$ and $T_{2}$, defining the CR-CC(2,3) approach \cite{crccl}. The main advantages of CC methods, as compared to truncated CI approaches, are size extensivity and the computationally efficient description of the higher-order components of the wave function beyond a CI model space truncated at the same excitation level via products of cluster operators. For example, the basic CCSD method describes $4p4h$ excitations through, e.g., a product of two $T_{2}$ clusters. The details of the CCSD and CR-CC(2,3) methods and their comparison to IT-CI can be found in Ref. \cite{RoGo09}.

In the CC approaches the evaluation of expectation values of operators is more involved than in the CI methods. As explained in Ref. \cite{RoGo09}, we have to rely on a response formulation or the equivalent derivatives of CC energies, since the traditional expectation value expression with the CC wave function produces a non-terminating power series in cluster amplitudes that does not lead to practical computational schemes. Moreover, non-iterative CC methods, such as CR-CC(2,3), rely on corrections to the energy only and do not have the corresponding wave function. Thus, to determine the CC analogs of $\expect{\HO_{\intr}}_{\beta}$ and $\expect{\HO_{\cm}}_{\beta}$, we proceed as follows. First, we evaluate the CC energy $E_{\beta^{\prime}}$ for the Hamiltonian $\HO_{\beta^{\prime}} = \HO_{\intr} + \beta^{\prime} \HO_{\cm}$ with $\beta^{\prime} = \beta - \Delta \beta$, $\beta$, and $\beta + \Delta \beta$ around the nominal value of $\beta$ we are interested in, where the step $\Delta \beta$ equals 0.01. Then, using a centered finite-difference form to approximate the first derivative, we compute $\expect{\HO_{\cm}}_{\beta}$ as $(\partial E_{\beta^{\prime}}/\partial {\beta^{\prime}})_{\beta^{\prime} = \beta}$ and from that obtain $\expect{\HO_{\intr}}_{\beta} = E_{\beta} - \beta \expect{\HO_{\cm}}_{\beta}$. We have checked the stability of the finite-difference calculations of $\expect{\HO_{\intr}}_{\beta}$ and $\expect{\HO_{\cm}}_{\beta}$ by including up to five $\beta^{\prime}$ points around $\beta$ and using a few different $\Delta \beta$ values. A cross-check of this prescription was also performed for the IT-CI approach, where a comparison with the exact expectation values was possible. The experience with performing CC calculations in quantum chemistry is that the difference between the conventional quantum-mechanical expectation values and the corresponding energy derivatives, as described above, are very small, since the approximate CC methods, such as those used in this work, provide results close to full CI and full CI, being an exact diagonalization, satisfies the Hellmann-Feynman theorem.

All calculations were performed with the $\VO_{\UCOM}$ interaction introduced in Refs. \cite{RoHe05,RoPa06} and employed in Refs. \cite{RoNa07,RoGo09,Roth09}. It is a pure two-body interaction derived from the Argonne V18 potential \cite{WiSt95} via a unitary transformation to account for short-range central and tensor correlations \cite{RoNe04}. We consider a few representative oscillator frequencies from the range $\hbar\Omega=22-38$ MeV, which includes the minima of the ground-state energy of \elem{O}{16} obtained with the many-body methods and single-particle basis sets used in this work (see Ref. \cite{RoGo09}).

\section{Illustration of the center-of-mass diagnostics}

As a first demonstration of the impact of the model space truncation on the coupling of intrinsic and CM motions, we analyze the IT-NCSM(seq) and IT-CI($4p4h$) results for the ground state of \elem{O}{16}. The IT-NCSM(seq) approach is based on the $N_{\max}\hbar\Omega$ space of full NCSM, which allows for the exact wave function factorization given by Eq. \eqref{eq:decoupling}, but the importance truncation formally breaks this factorization property and so we have to check the magnitude of the resulting CM contamination. In the case of IT-CI($4p4h$), the underlying CISDTQ model space violates translational invariance even before the importance truncation is invoked and so we expect the unphysical coupling between intrinsic and CM degrees of freedom to be more severe than in the IT-NCSM(seq) case.

\begin{figure}
\centering\includegraphics[width=0.7\columnwidth]{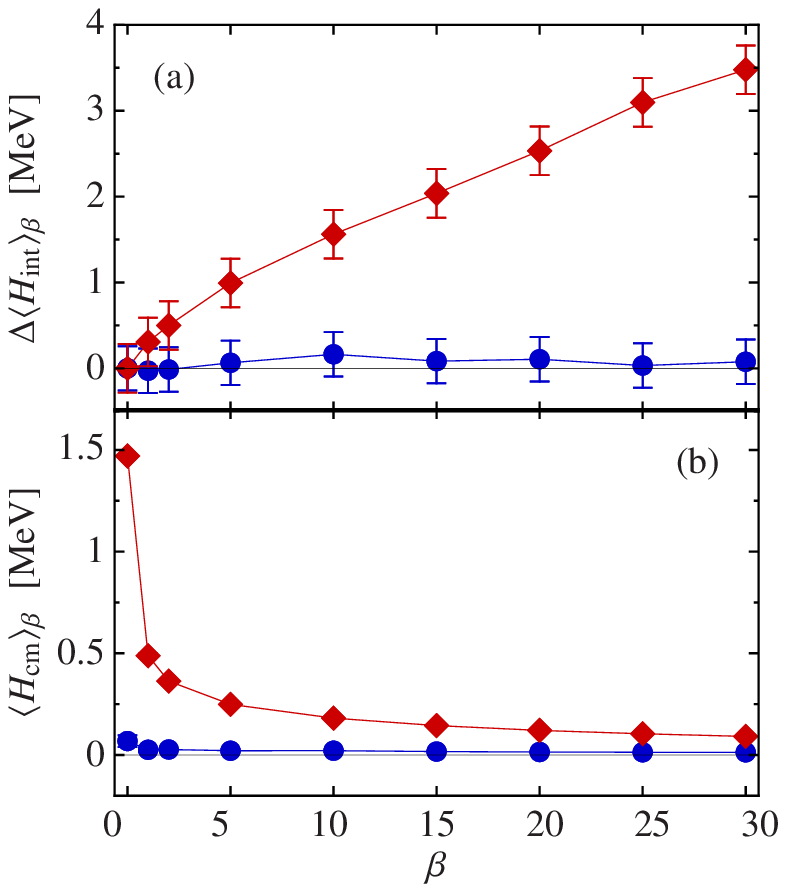}
\caption{Dependence of the intrinsic energy change $\Delta\expect{\HO_{\intr}}_{\beta}$ (a) and of the expectation value $\expect{\HO_{\cm}}_{\beta}$ (b) on $\beta$ in IT-NCSM(seq) (\symbolcircle[FGBlue]) and IT-CI($4p4h$) (\symboldiamond[FGRed]) calculations for the ground state of \elem{O}{16}. In IT-CI($4p4h$), $e_{\max}=5$ and $\hbar\Omega=30$ MeV. In IT-NCSM(seq), $N_{\max}=8$ and $\hbar\Omega=30$ MeV.}
\label{fig:betadep}
\end{figure}

We study the dependence of the expectation values of the intrinsic Hamiltonian, $\expect{\HO_{\intr}}_{\beta}$, and of the CM Hamiltonian, $\expect{\HO_{\cm}}_{\beta}$, on the parameter $\beta$ entering $\HO_{\beta}$, Eq. \eqref{eq:hamiltonianmod}. Figure \ref{fig:betadep} shows this dependence for an IT-NCSM(seq) calculation with $N_{\max}=8$ and for an IT-CI($4p4h$) calculation with $e_{\max}=5$, both at fixed $\hbar\Omega=30$ MeV. For IT-NCSM(seq), the change of the intrinsic energy $\Delta\expect{\HO_{\intr}}_{\beta} = \expect{\HO_{\intr}}_{\beta} - \expect{\HO_{\intr}}_{0}$ is always below $150$ keV. Considering the error bars resulting from the threshold extrapolation, this is consistent with the exact NCSM value of zero. The lack of dependence of $\expect{\HO_{\intr}}_{\beta}$ on $\beta$ shows that the IT-NCSM(seq) eigenstates satisfy the wave function factorization given by Eq. (\ref{eq:decoupling}) to a very good approximation. In contrast, the IT-CI($4p4h$) calculations show a sizable change in $\expect{\HO_{\intr}}_{\beta}$, of up to $3$ MeV when $\beta$ increases to 30, which is a signature of the significant coupling of the intrinsic and CM motions in the IT-CI($4p4h$) wave function.

The expectation values of the CM Hamiltonian, $\expect{\HO_{\cm}}_{\beta}$, exhibit complementary patterns. In the case of the IT-NCSM(seq) calculations, they are at the level of $70$ keV for $\beta=0$ and drop to below $20$ keV already for $\beta=1$. In contrast, the IT-CI($4p4h$) calculations result in the $\expect{\HO_{\cm}}_{\beta}$ values of about 1.5 MeV for $\beta=0$ which are suppressed to $0.5$ MeV and $0.2$ MeV for $\beta=1$ and $10$, respectively. One might think that an expectation value of $\HO_{\cm}$ of 200 keV is sufficiently small to indicate a CM decoupling. However, despite the smallness of $\expect{\HO_{\cm}}_{\beta}$ the intrinsic energy continues to change, by $\sim 1$ MeV when going from $\beta=10$ to $\beta=20$. This shows that a value of $\expect{\HO_{\cm}}_{\beta}$ of a few hundred keV is not sufficient to claim that intrinsic and CM motions decouple.

The general $\beta$-dependence of the IT-CI(4p4h) results shown in Fig. 1 also holds for the CC calculations. Keeping this systematics in mind, we simplify further discussion and consider $\expect{\HO_{\intr}}_{\beta}$ and $\expect{\HO_{\cm}}_{\beta}$ for $\beta=0$ and $10$ only, investigating their behavior as functions of mass number, model-space size, and oscillator frequency.

\section{Case Study: \elem{He}{4}}

As a first case, we study the systematics of the CM contamination in the IT-CI($4p4h$)\footnote{For IT-CI($4p4h$) the uncertainties due to the threshold extrapolation are below $0.05$ MeV for $\expect{\HO_{\intr}}$ and below $0.01$ MeV for $\expect{\HO_{\cm}}$.}, CCSD, and CR-CC(2,3) calculations for the ground state of \elem{He}{4}. Table \ref{tab:he4} summarizes the expectation values of $\HO_{\intr}$ and $\HO_{\cm}$ obtained for $\beta=0$ and $10$ for different $e_{\max}$-truncated model spaces and different oscillator frequencies $\hbar\Omega$. Instead of the intrinsic energy at $\beta=10$, we list the energy change $\Delta\expect{\HO_{\intr}}_{\beta=10}$. 

\begin{table}[t]
\caption{Center-of-mass diagnostics for the ground state of \elem{He}{4} using the $\VO_{\UCOM}$ interaction. All energies are in units of MeV.}
\label{tab:he4}
\small
\begin{tabular}{l c c c c c c}
\hline
& & & \multicolumn{2}{c}{$\beta=0$} & \multicolumn{2}{c}{$\beta=10$} \\
Method & $\hbar\Omega$ & $e_{\max}$ & $\expect{\HO_{\intr}}$ & $\expect{\HO_{\cm}}$ & $\Delta\expect{\HO_{\intr}}$ & $\expect{\HO_{\cm}}$ \\
\hline 
IT-CI($4p4h$) & 30 & 4 & -25.992 & 3.638 & 0.568 & 0.027 \\
         &    & 5 & -26.809 & 1.149 & 0.311 & 0.028 \\
         &    & 6 & -27.412 & 2.524 & 0.190 & 0.017 \\
         &    & 7 & -27.758 & 1.978 & 0.113 & 0.017 \\
         &    & 8 & -28.021 & 1.799 & 0.121 & 0.017 \\
         & 38 & 4 & -26.313 & 3.372 & 0.641 & 0.030 \\
         &    & 5 & -27.184 & 0.911 & 0.464 & 0.040 \\
         &    & 6 & -27.777 & 3.234 & 0.203 & 0.020 \\
         &    & 7 & -28.055 & 2.612 & 0.213 & 0.021 \\
         &    & 8 & -28.192 & 3.254 & 0.159 & 0.019 \\ 
\hline 
CCSD     & 30 & 4 & -25.537 & 3.639 & 0.699 & 0.038 \\
         &    & 5 & -26.319 & 2.465 & 0.585 & 0.027\\
         &    & 6 & -26.887 & 2.976 & 0.493 & 0.024 \\
         & 38 & 4 & -25.679 & 8.158 & 1.069 & 0.046\\
         &    & 5 & -26.413 & 6.346 & 0.952 & 0.039\\
         &    & 6 & -27.035 & 8.965 & 0.924 & 0.023 \\
\hline 
CR-CC(2,3)&30 & 4 & -25.995 & 4.049 & 0.694 & 0.047 \\
         &    & 5 & -26.867 & 2.291 & 0.605 & 0.042 \\
         &    & 6 & -27.536 & 3.347 & 0.556 & 0.039  \\
         & 38 & 4 & -26.390 & 6.282 & 0.901 & 0.070 \\
         &    & 5 & -27.261 & 3.575 & 0.810 & 0.073 \\
         &    & 6 & -27.975 & 7.641 & 0.766 & 0.058 \\
\hline
\end{tabular}
\end{table}%

Let us first consider the expectation values of $\HO_{\cm}$ for $\beta=0$. In the case of IT-CI($4p4h$), $\expect{\HO_{\cm}}_{\beta=0}$ ranges from $0.9$ to $3.6$ MeV without a clear trend with respect to model-space size. For CCSD and CR-CC(2,3), the $\expect{\HO_{\cm}}_{\beta=0}$ values range from $2.5$ to $9$ MeV and $2.3$ to $7.6$ MeV, respectively. These are large values when compared to the intrinsic energies. If one used $\expect{\HO_{\cm}}_{\beta=0}$ as the sole criterion, as has been done in the CC context (cf., e.g., \cite{HaPa09}), one would have to conclude that the extent of CM contamination is dramatic. However, when considering the results for $\beta=10$ a different picture emerges. For $e_{\max}=6$, the change in the intrinsic energy $\Delta\expect{\HO_{\intr}}_{\beta=10}$ is on the order of $200$ keV for IT-CI($4p4h$), $900$ keV for CCSD, and $800$ keV for CR-CC(2,3), implying a sizable CM contamination but not as severe as indicated by $\expect{\HO_{\cm}}_{\beta=0}$. Furthermore, $\Delta\expect{\HO_{\intr}}_{\beta=10}$ decreases with increasing $e_{\max}$. This behavior is confirmed by the values of $\expect{\HO_{\cm}}_{\beta=10}$, which are on the order of a few tens of keV and decrease with increasing $e_{\max}$ as well. Both quantities indicate that CM contaminations are larger for $\hbar\Omega=38$ MeV than for $\hbar\Omega=30$ MeV, but they are not as large as one might expect based on the fact that \elem{He}{4} is a light nucleus.

The above observations can be explained in the following manner. The exact solution of the Schr{\" o}dinger equation, which leads to perfect decoupling of intrinsic and CM degrees of freedom, is approached when the CI or CC calculation allows for all possible $npnh$ excitations and when $e_{\max}$ approaches $\infty$. \elem{He}{4} consists of only four particles, so IT-CI($4p4h$) becomes an exact theory when the importance truncation threshold vanishes and $e_{\max}\to\infty$. The relatively small values of $\Delta\expect{\HO_{\intr}}_{\beta=10}$ and $\expect{\HO_{\cm}}_{\beta=10}$ and their systematic reduction with increasing $e_{\max}$ are, therefore, expected and this is what we observe in our IT-CI($4p4h$) calculations in which we extrapolate the results to the CISDTQ limit. When an issue of size extensivity is of no importance, as is the case for a four-particle \elem{He}{4} problem, the CR-CR(2,3) approach represents a very good approximation to IT-CI($4p4h$) or CISDTQ \cite{RoGo09} and thus the systematic reduction of $\Delta\expect{\HO_{\intr}}_{\beta=10}$ and $\expect{\HO_{\cm}}_{\beta=10}$ with increasing $e_{\max}$ and their relatively small values, though not on the same level as in the IT-CI($4p4h$) case, can be expected and our calculations confirm this.

We can draw three conclusions from the \elem{He}{4} results. First, the magnitude of $\expect{\HO_{\cm}}_{\beta=0}$ is not a viable criterion for assessing the extent of the CM decoupling; in some cases, the $\expect{\HO_{\cm}}_{\beta=0}$ values do not even reproduce the qualitative trends. Second, the change of the intrinsic energy $\Delta\expect{\HO_{\intr}}_{\beta=10}$ when going from $\beta=0$ to $10$ provides a direct and robust measure of the degree of CM contamination and its effect on the intrinsic energy, with $\expect{\HO_{\cm}}_{\beta=10}$ offering additional insights. Third, the $\Delta\expect{\HO_{\intr}}_{\beta}$ and $\expect{\HO_{\cm}}_{\beta}$ values for nonzero $\beta$ show a systematic reduction of the CM contamination as $e_{\max}\to\infty$. This makes sense since in the $e_{\max}\to\infty$ limit IT-CI($4p4h$) and CR-CC(2,3) essentially recover the full Hilbert space of the four-particle system and thus effectively approach the exact and factorized solution of the Schr{\" o}dinger equation.

\section{Case Study: \elem{O}{16}}

In order to examine the CM contamination in the CI and CC calculations for heavier nuclei, we analyze \elem{O}{16}. A common presumption is that CM contaminations are less severe as $A$ increases because of a general $1/A$-scaling contained in the CM Hamiltonian, Eq. \eqref{eq:cmhamiltonian}.

\begin{table}[t]
\caption{Center-of-mass diagnostics for the ground state of \elem{O}{16} using the $\VO_{\UCOM}$ interaction. All energies are in units of MeV.}
\label{tab:o16}
\small
\begin{tabular}{c c c c c c c}
\hline
& & & \multicolumn{2}{c}{$\beta=0$} & \multicolumn{2}{c}{$\beta=10$} \\
Method & $\hbar\Omega$ & $e_{\max}$ & $\expect{\HO_{\intr}}$ & $\expect{\HO_{\cm}}$ & $\Delta\expect{\HO_{\intr}}$ & $\expect{\HO_{\cm}}$ \\
\hline 
IT-CI($4p4h$) 
         & 22 & 4 &  -94.80 & 0.45 & 0.93 & 0.10 \\
         &    & 5 & -103.62 & 0.51 & 1.03 & 0.10 \\
         &    & 6 & -110.14 & 0.37 & 1.57 & 0.09 \\
         &    & 7 & -115.03 & 0.37 & 1.92 & 0.08 \\
         & 30 & 4 &  -87.65 & 0.81 & 1.28 & 0.18 \\
         &    & 5 &  -98.67 & 1.47 & 1.56 & 0.18 \\
         &    & 6 & -104.24 & 1.25 & 1.84 & 0.16 \\
         &    & 7 & -108.43 & 1.30 & 2.17 & 0.15 \\
         & 38 & 4 & -58.62  & 1.24 & 2.43 & 0.34 \\
         &    & 5 & -74.75  & 2.61 & 1.87 & 0.35 \\
         &    & 6 & -79.52  & 2.43 & 1.49 & 0.32 \\
         &    & 7 & -83.72  & 2.73 & 2.58 & 0.31 \\
         &    & 8 & -85.81  & 2.56 & 2.41 & 0.30 \\
         &    & 9 & -88.81  & 2.62 & 3.21 & 0.28 \\
\hline 
CCSD     & 22 & 4 &  -94.79 & 1.11 & 1.13 & 0.18 \\
         &    & 5 & -103.94 & 2.16 & 1.29 & 0.18 \\
         &    & 6 & -109.97 & 1.68 & 1.09 & 0.14 \\
         &    & 7 & -114.73 & 1.80 & & \\
         & 30 & 4 &  -93.01 & 1.08 & 1.76 & 0.26 \\
         &    & 5 & -107.32 & 5.88 & 2.49 & 0.24 \\
         &    & 6 & -113.06 & 5.31 & 2.37 & 0.20 \\
         &    & 7 & -118.15 & 7.16 & & \\
         & 38 & 4 &  -77.09 & 3.78 & 3.65 & 0.44 \\
         &    & 5 & -102.85 & 8.83 & 3.30 & 0.29 \\
         &    & 6 & -109.28 & 8.21 & 3.31 & 0.28 \\
         &    & 7 & -116.86 & 15.20  & &\\
\hline 
CR-CC(2,3)&22 & 4 &  -98.10 & 1.06 & 1.30 & 0.10 \\
         &    & 5 & -108.12 & 2.60 & 1.45 & 0.11\\
         &    & 6 & -114.81 & 1.96 & 1.24 & 0.13\\
         &    & 7 & -120.21 & 2.29 & & \\
         & 30 & 4 &  -97.78 & 0.62 & 2.15 & 0.16 \\
         &    & 5 & -113.14 & 5.38 & 1.91 & 0.20 \\
         &    & 6 & -119.92 & 4.62 & 1.67 & 0.24 \\
         &    & 7 & -125.92 & 7.45 & & \\
         & 38 & 4 &  -84.16 & 10.74 & 5.63 & 0.26 \\
         &    & 5 & -109.77 &  6.46 & 2.10 & 0.38 \\
         &    & 6 & -117.62 &  5.09 & 1.70 & 0.47 \\
         &    & 7 & -126.16 & 14.51 & & \\
\hline
\end{tabular}
\end{table}

The IT-CI($4p4h$)\footnote{For IT-CI($4p4h$) the extrapolation uncertainties are $0.2 - 0.4$ MeV for $\expect{\HO_{\intr}}$ and below $0.01$ MeV for $\expect{\HO_{\cm}}$.}, CCSD, and CR-CC(2,3) results are summarized in Table \ref{tab:o16}. We first focus on $\hbar\Omega=38$ MeV, which yields the minimum energy in the CR-CC(2,3) calculation for $e_{\max}=7$ \cite{RoGo09}, the largest model space we are able to handle in the CC case. The expectation values of $\HO_{\cm}$ for $\beta=0$ are very large, reaching about $3$ MeV for IT-CI($4p4h$) and about $15$ MeV for CCSD and CR-CC(2,3), with no clear trend as functions of $e_{\max}$. 

For a quantitative assessment we again resort to calculations using $\HO_{\beta}$ with $\beta=10$. When $\hbar\Omega=38$ MeV, the typical $\Delta\expect{\HO_{\intr}}_{\beta=10}$ values in the larger model spaces are about $2.5$ MeV for IT-CI($4p4h$), $3$ MeV for CCSD, and $2$ MeV for CR-CC(2,3), i.e., all schemes suffer from a sizable CM contamination of the intrinsic energies. The energy changes $\Delta\expect{\HO_{\intr}}_{\beta=10}$ are rather stable and do not show a clear trend for the different model spaces considered here, except for CR-CC(2,3) which seems to show a decrease with increasing $e_{\max}$. The expectation values of $\HO_{\cm}$ for $\beta=10$ and $\hbar\Omega=38$ MeV are in the $250-500$ keV range. For IT-CI($4p4h$) and CCSD they decrease with increasing $e_{\max}$, whereas for CR-CC(2,3) they slightly increase. Due to this behavior and our computational limitations we are unable to draw definite conclusions on a possible reduction of the CM contamination when going to model spaces with very large $e_{\max}$. Note, however, that unlike in the \elem{He}{4} case, neither IT-CI($4p4h$) nor CCSD or CR-CC(2,3) recover the full many-body Hilbert space for \elem{O}{16} in the $e_{\max}\to\infty$ limit, because none of these approaches describes all of the relevant $npnh$ excitations. Therefore, even when $e_{\max}\to\infty$, the factorization of the resulting ground states is not possible.

Concerning the dependence on $\hbar\Omega$, we observe a systematic increase of $\Delta\expect{\HO_{\intr}}_{\beta=10}$ and $\expect{\HO_{\cm}}_{\beta=10}$ with increasing $\hbar\Omega$ for all CI and CC methods and all model-space sizes used in this work. Although $\expect{\HO_{\cm}}_{\beta=0}$ follows this trend as well, a closer inspection of the behavior of $\expect{\HO_{\cm}}_{\beta=0}$ in comparison to $\Delta\expect{\HO_{\intr}}_{\beta=10}$ again reveals that the expectation value of $\HO_{\cm}$ at $\beta=0$ cannot be used to quantify the degree of CM contamination. For example, in the IT-CI($4p4h$) calculations for $\hbar\Omega=22$ MeV and $e_{\max}=7$ we obtain $\expect{\HO_{\cm}}_{\beta=0}$ of less than $400$ keV although $\Delta\expect{\HO_{\intr}}_{\beta=10}$ is about $1.9$ MeV. We find the virtually identical $\Delta\expect{\HO_{\intr}}_{\beta=10}$ value in the CR-CC(2,3) calculations for $\hbar\Omega=30$ MeV and $e_{\max}=5$, but the corresponding $\expect{\HO_{\cm}}_{\beta=0}$ of $5.38$ MeV is larger than the aforementioned IT-CI($4p4h$) value of $\expect{\HO_{\cm}}_{\beta=0}$ by more than an order of magnitude.

If we compare the general picture emerging from the IT-CI($4p4h$), CCSD, and CR-CC(2,3) calculations for \elem{O}{16}, where $\Delta\expect{\HO_{\intr}}_{\beta=10}$ is typically in the range of $1-3$ MeV, with the analogous results for \elem{He}{4}, where $\Delta\expect{\HO_{\intr}}_{\beta=10}$ is typically $100 - 900$ keV, then the popular belief that the CM contamination is suppressed as $1/A$ becomes questionable. There are other mechanisms affecting the CM contamination which counteract a simplistic $1/A$ scaling, the truncation of the many-body model space at a fixed excitation level $n < A$ and the accuracy of a given many-body method relative to full CI, which accounts for all $npnh$ excitations, being some of them. The restriction of the model space to up to $4p4h$ states in IT-CI($4p4h$) provides a far less complete approximation to the full Hilbert space in the case of \elem{O}{16} than in the \elem{He}{4} case, where IT-CI($4p4h$) is virtually exact. For CCSD and CR-CC(2,3) the situation is more complex, since the truncation of the cluster operator $T$, which enters the exponential ansatz for the wave function, at $T_{2}$ or $T_{3}$ brings a variety of product excitations through the {\it disconnected} clusters, such as $(1/2) T_{1} T_{2}^{2}$, $(1/6) T_{2}^{3}$, etc., which are beyond the $4p4h$ excitation level of CI, but still, CCSD and CR-CC(2,3) are not the exact theories, so the problem remains. For the closed-shell nuclei described by pairwise interactions, such as \elem{O}{16}, the role of higher-order {\it connected} clusters, such as, e.g., $T_{4}$, in describing the intrinsic energy, is expected to be small \cite{ni56_2007}, but it is not entirely clear what the significance of the higher-order cluster components of the wave function neglected in CCSD and CR-CC(2,3) on the CM motion and its coupling with the intrinsic state of the nucleus is. Therefore, the ability of the model space used in the CI and CC calculations to represent a factorized eigenstate $|\Psi\rangle$ is getting progressively worse with increasing $A$. 

\section{Conclusions}

We have formulated simple and rigorous criteria for diagnosing the CM contamination in many-body approaches to the nuclear structure problem. They are based on the fact that for a factorized many-body state the CM component can be manipulated, e.g., by adding a CM contribution $\beta \HO_{\cm}$ to the intrinsic Hamiltonian $\HO_{\intr}$,  without affecting the intrinsic part. The existence of an unphysical coupling between CM and intrinsic motions manifests itself in a dependence of the change of the intrinsic energy expectation value $\Delta\expect{\HO_{\intr}}_{\beta}$ on $\beta$, since $\Delta\expect{\HO_{\intr}}_{\beta} = 0$ for all $\beta$ when the nuclear state factorizes. The analysis of $\Delta\expect{\HO_{\intr}}_{\beta}$ is our primary diagnostic. A secondary measure of the magnitude of the coupling between CM and intrinsic motions is the expectation value of $\HO_{\cm}$ at non-zero $\beta$, which should be zero for factorizable states. In contrast to these two quantities, the expectation value $\expect{\HO_{\cm}}_{\beta}$ at $\beta=0$, which is sometimes used to examine the degree of CM contamination, does not provide a meaningful measure since it can have any positive value for a many-body state that factorizes. Furthermore, we have shown examples where $\expect{\HO_{\cm}}_{\beta=0}$ is on the order of a few hundred keV when $\Delta\expect{\HO_{\intr}}_{\beta=10}$ is on the order of a few MeV, indicating a sizable CM contamination.

Using our diagnostics we have demonstrated that the IT-NCSM approach, which is based on the $N_{\max}\hbar\Omega$ model-space cutoff, exhibits a very good approximate factorization with $\Delta\expect{\HO_{\intr}}_{\beta}$ below 150 keV for all $\beta$ in spite of the fact that the importance truncation used to select the dominant contributions to the NCSM wave function formally breaks the rigorous factorization of the $N_{\max}\hbar\Omega$ space. Switching from the $N_{\max}$- to the $e_{\max}$-truncation of IT-CI($4p4h$) leads to a sizable CM contamination, with $\Delta\expect{\HO_{\intr}}_{\beta}$ reaching a few MeV as $\beta$ increases. Similarly, the CCSD and CR-CC(2,3) calculations show sizable CM contaminations, with $\Delta\expect{\HO_{\intr}}_{\beta=10}$ typically somewhat larger than in the IT-CI($4p4h$) case. With increasing $e_{\max}$, the degree of CM contamination decreases for \elem{He}{4}, but one does not observe the same clear trend for \elem{O}{16}. This difference can be explained by the fixed maximum excitation level of the truncated CI and CC calculations used in this work. The exact solution of the Schr{\" o}dinger equation, which leads to perfect decoupling of intrinsic and CM degrees of freedom, is approached only when the CI or CC calculation allows for all possible $npnh$ excitations and when $e_{\max}\rightarrow\infty$. The fixed excitation level $n < A$ used in the CI and CC calculations for \elem{O}{16} prevents this.

Our findings seem to disagree with a recent claim of Hagen \emph{et al.} \cite{HaPa09} of a factorization of the CC ground states to a very good approximation, based on the observation that the expectation value $\expect{\tilde{\HO}_{\cm}}_{\beta=0}$ for a generalized operator $\tilde{\HO}_{\cm}$ defined for an oscillator frequency $\tilde{\Omega}\neq\Omega$ minimizing $\expect{\tilde{\HO}_{\cm}}_{\beta=0}$ is close to zero. For \elem{O}{16} described by the N$^3$LO interaction \cite{EnMa03} in a Hartree-Fock basis with $e_{\max}\approx18$ they obtain the optimum $\expect{\tilde{\HO}_{\cm}}_{\beta=0}$ values in the range of about $-0.5$ to $1$ MeV when varying $\hbar\Omega$ from $24$ to $52$ MeV. Aside from concerns about the interpretation of negative expectation values for a positive definite operator and the validity of the generalization of $\tilde{\HO}_{\cm}$, the values of $\expect{\tilde{\HO}_{\cm}}_{\beta=0}$ observed in \cite{HaPa09} are not sufficient to claim the approximate factorization of the CC ground states. In our view, it is more appropriate to monitor $\Delta\expect{\HO_{\intr}}_{\beta}$ and $\expect{\HO_{\cm}}_{\beta}$ at finite $\beta$ to assess the degree of coupling of intrinsic and CM degrees of freedom in truncated CI and CC calculations.

\section*{Acknowledgments}

We thank Petr Navr\'atil for discussions. This work was supported by the DFG through SFB 634, the Helmholtz International Center for FAIR (R.R.), the U.S. Department of Energy (Grant No. DE-FG02-01ER15228; P.P), and the National Science Foundation's Graduate Research Fellowship (J.R.G).



\begin{thebibliography}{25}

\bibitem{NaKa00}
P. Navr\'atil, G. P. Kamuntavicius, B. R. Barrett, 
Phys. Rev. C 61 (2000) 044001. 

\bibitem{NaQu09}
P. Navr\'atil, S. Quaglioni, I. Stetcu, B. R. Barrett,
J. Phys. G: Nucl. Part. Phys. 36 (2009) 083101.

\bibitem{BaLe99}
N. Barnea, W. Leidemann, G. Orlandini, 
Nucl. Phys. A650 (1999) 427.  
  
\bibitem{PiWi01}
S.~C. Pieper, R.~B. Wiringa,
Ann. Rev. Nucl. Part. Sci. 51 (2001) 53.

\bibitem{Palu67}
F. Palumbo, 
Nucl. Phys. A99 (1967) 100.

\bibitem{GlLa74}
D. Gloeckner, R. Lawson, 
Phys. Lett. 53B (1974) 313.

\bibitem{RaFa90}
P. K. Rath, A. Faessler, H. M\"uther, A. Watts, 
J. Phys. G: Nucl. Part. Phys. 16 (1990) 245.

\bibitem{RoNa07}
R. Roth, P. Navr\'atil, 
Phys. Rev. Lett. 99 (2007) 092501.

\bibitem{RoGo09}
R. Roth, J. R. Gour, P. Piecuch, 
Phys. Rev. C 79 (2009) 054325.

\bibitem{Roth09}
R. Roth, 
Phys. Rev. C 79 (2009) 064324. 

\bibitem{coester}
F. Coester, 
Nucl. Phys. 7 (1958) 421. 

\bibitem{coester2}
F. Coester, H. K\"ummel, 
Nucl. Phys. 17 (1960) 477.

\bibitem{cizek}
J. {\v C}{\'\i}{\v z}ek,
J. Chem. Phys. 45 (1966) 4256. 

\bibitem{cizek2}
J. {\v C}{\'\i}{\v z}ek,
Adv. Chem. Phys. 14 (1969) 35. 

\bibitem{ccsd}
G. D. Purvis III, R. J. Bartlett, 
J. Chem. Phys. 76 (1982) 1910.

\bibitem{crccl}
P. Piecuch, M. W{\l}och, 
J. Chem. Phys. 123 (2005) 224105.

\bibitem{kowalski04}
K. Kowalski \emph{et al.},
Phys. Rev. Lett. 92 (2004) 132501.

\bibitem{o16prl}
M. W{\l}och \emph{et al.},
Phys. Rev. Lett. 94 (2005) 212501. 
  
\bibitem{ni56_2007}
M. Horoi \emph{et al.},
Phys. Rev. Lett. 98 (2007) 112501.
  
\bibitem{RoHe05}
R. Roth \emph{et al.},
Phys. Rev. C 72 (2005) 034002.
  
\bibitem{RoPa06}
R. Roth \emph{et al.},
Phys. Rev. C 73 (2006) 044312.

\bibitem{WiSt95}
R. B. Wiringa, V. G. J. Stoks, R. Schiavilla, 
Phys. Rev. C 51 (1995) 38.

\bibitem{RoNe04}
R. Roth, T. Neff, H. Hergert, H. Feldmeier, 
Nucl. Phys. A745 (2004) 3.

\bibitem{HaPa09}
G. Hagen, T. Papenbrock, D. J. Dean (2009), 
arXiv:0905.3167.

\bibitem{EnMa03}
D. R. Entem, R. Machleidt, Phys. Rev. C 68 (2003) 041001(R).

\end{thebibliography}
\end{document}